\documentclass[prd, aps, superscriptaddress, preprintnumbers, floatfix, nofootinbib]{revtex4}

\usepackage{amsfonts}
\usepackage{amsmath}
\usepackage{amssymb}
\usepackage{bm}
\usepackage{dcolumn}
\usepackage{graphicx}   
\usepackage[latin1]{inputenc}
\usepackage{latexsym}
\usepackage{rotating}
\usepackage{graphicx}
\usepackage{color}
\usepackage{ulem}

\usepackage{amsfonts}
\usepackage{amsmath}
\usepackage{amssymb}
\usepackage{bm}
\usepackage{dcolumn}
\usepackage{epsfig}
\usepackage{graphicx}
\usepackage[latin1]{inputenc}
\usepackage{latexsym}
\usepackage{rotating}

\usepackage[unicode=true,pdfusetitle,
 bookmarks=true,bookmarksnumbered=false,bookmarksopen=false,
 breaklinks=false,pdfborder={0 0 1},backref=false,colorlinks=true]
 {hyperref}
\hypersetup{
 linkcolor=blue, citecolor=magenta, urlcolor=red, filecolor=blue}

\newcommand\be{\begin{equation}}
\newcommand\ba{\begin{eqnarray}}
\newcommand\ee{\end{equation}}
\newcommand\ea{\end{eqnarray}}

\newcommand{\llan}{\left\langle}
\newcommand{\rran}{\right\rangle}
\newcommand{\sfst}[1]{^{#1}}

\def\snd{^{(2)}}

\def\conh{\mathcal{H}}

\begin{document}

\title {Back-Reaction of Gravitational Waves Revisited}

\author{Robert Brandenberger}
\email{rhb@physics.mcgill.ca}
\affiliation{Physics Department, McGill University, Montreal, QC, H3A 2T8, Canada}

\author{Tomo Takahashi}
\email{tomot@cc.saga-u.ac.jp}
\affiliation{Department of Physics, Saga University, Saga 840-8502, Japan}

\date{\today}

\begin{abstract}

We study the back-reaction of gravitational waves in early universe cosmology, focusing
both on super-Hubble and sub-Hubble modes. Sub-Hubble modes lead to an effective
energy density which scales as radiation. Hence, the relative contribution of such gravitational waves
to the total energy density is constrained by big bang nucleosynthesis. This leads to
an upper bound on the tensor spectral slope $n_T$ which also depends on the
tensor to scalar ratio $r$. Super-Hubble modes, on the other hand, lead to a
negative contribution to the effective energy density, and to an equation of state
of curvature. Demanding that the early universe is not dominated by the back-reaction
leads to constraints on the gravitational wave spectral parameters which are derived.

\end{abstract}


\maketitle

\section{Introduction} 

Gravitational waves (GW) carry energy and momentum and hence affect the background space-time
they propagate in. This is a basic consequence of the nonlinearity of the Einstein field equations.
The effects of gravitational waves propagating on a given space-time on the space-time itself
is called {\it gravitational back-reaction}.

A spectrum of primordial gravitational waves is generated \cite{Starob} in early universe scenarios 
such as inflationary cosmology \cite{infl}, but also \cite{BNPV2} in alternatives such as string gas 
cosmology \cite{BV, SGrev}. The spectrum is characterized by its amplitude ${\cal A}_T$ and 
spectral tilt $n_T$. If $k_{*}$ is the comoving wavenumber used as the pivot scale (e.g. the
scale of the quadrupole of the cosmic microwave background (CMB)), then the power spectrum
of gravitational waves on scale $k$ is given by
\be
{\cal P}_T(k) \, = \,  A_T(k_{*}) \left( \frac{k}{k_{*}} \right)^{n_T}  \, .
\ee
The amplitude of the gravitational wave spectrum at the pivot scale is related
to the observed amplitude of the spectrum of cosmological perturbations on that
scale via the tensor-to-scalar ratio $r_{*}$, with the subscript $*$ denoting that the
amplitudes are evaluated at the pivot scale. The current upper bound on the value
of $r_{*}$ is $0.07$ at 95 \% C.L.~\cite{BICEP}. Even assuming that the amplitude of gravitational
waves is close to the current upper bound, the current limits on the slope $n_T$ are weak
\cite{Stewart}. It is interesting to explore the possibility that gravitational back-reaction
considerations can lead to improved constraints.

Short wavelength (wavelengths smaller than the Hubble radius) gravitational waves
oscillate and have the same equation of state as radiation. Hence, Big Bang 
Nucleosynthesis sets an upper bound on the energy in short wavelength gravitational waves.

The effects of long wavelength (super-Hubble) waves is much less studied. It is
sometimes claimed that causality prevents such modes from having a locally
measurable effect (see, e.g., \cite{Wald}). This, however, is incorrect. First of all,
in all models which yield a solution of the {\it horizon problem} of Standard Big Bang
Cosmology, and in which a causal explanation of the origin of structure is possible,
the cosmological horizon (the forward light cone of a point at the time when the
initial conditions are imposed) is much larger than the Hubble radius. In the
case of inflationary cosmology, the horizon at the end of the period of inflation has
a radius which is larger by a factor of $e^{N}$ than the Hubble radius, where $N$
is the total number of $e$-foldings of inflation. Hence there is no causality argument
which prohibits super-Hubble (but sub-horizon) modes from having a back-reaction
effect (see e.g.~\cite{Fabio1} for a discussion of this point). In fact, since super-Hubble
gravitational waves give a non-vanishing contribution to the local effective
energy-momentum tensor, they are able to have a local effect on the background
geometry.

There has been more focus on the back-reaction effects of super-Hubble
scalar cosmological perturbations (scalar fluctuations about a homogeneous
and isotropic background). If we consider linearized cosmological fluctuations,
then the effective energy-momentum tensor with which these fluctuations back-react
on the background metric is quadratic in the amplitude of the fluctuations. In
\cite{Abramo1, Abramo2} the form of this effective energy-momentum tensor was
derived, and it was shown that the effective energy-momentum tensor of
super-Hubble modes has the form of a negative cosmological constant. 
This follows from the fact that both spatial and temporal derivative terms in
the effective energy-momentum tensor are suppressed on super-Hubble scales,
leaving only terms which act like scalar field potential energy. Since on super-Hubble
scales the negativity of the effective gravitational energy overwhelms the positivity
of the matter energy, the contribution is like that of a negative cosmological
constant. This gives rise to the possibility of a dynamical relaxation mechanism
for the cosmological constant (see \cite{RHBbrReview} for a review and
\cite{Polyakov, WT1} for similar ideas).

The effects of super-Hubble gravitational waves in a de Sitter universe (no matter
present) has been considered in a series of papers by Tsamis and Woodard \cite{WT}. They
find that at two loop order, long wavelength modes induce a negative contribution
to the cosmological constant and could thus lead to a relaxation of the bare
cosmological constant. On the other hand, in \cite{Abramo2} it was shown that
in a universe which contains matter and hence scalar metric perturbations, the
super-Hubble modes have an equation of state (mode by mode)
\be
p \, = \, - \frac{1}{3} \rho \, ,
\ee
with $p$ and $\rho$ denoting pressure and energy densities, respectively. This is
the equation of state of spatial curvature. As for scalar metric fluctuations, the
sign of the effective energy density is negative.

As first pointed out by Unruh \cite{Unruh}, the local observability of back-reaction
effects of super-Hubble modes is an important issue. In fact, it was shown
in \cite{Ghazal1, AW} that for adiabatic cosmological perturbations the back-reaction
effect of super-Hubble modes is equivalent to a second order time translation.
On the other hand, in models with a separate clock field (like the radiation field
in our current matter-dominated cosmology) the back-reaction effects of
super-Hubble modes are locally measurable \cite{Ghazal2}, and they 
correspond to a decrease in the observed Hubble expansion rate \cite{Vacca}. 
Locally, the back-reaction of infrared modes manifests itself in a change of
the locally measured cosmological constant and curvature scalar \cite{Lam}.

In a model which contains matter, the physical time $t$ which appears in
the Friedmann-Robertson-Walker line element (see below) is related to
the energy density of matter. Matter plays the role of a physical clock,
and hence any effects on the local Hubble expansion rate which we find
as a consequence of including long wavelength gravitational waves is
a physical effect and not a gauge artefact.

In this paper we revisit the question of back-reaction effects of gravitational
waves in a universe which is dominated by its matter content. We consider
the back-reaction effects of the totality of the modes,
both infrared (super-Hubble) and ultraviolet (sub-Hubble). We study
not only a nearly de Sitter phase of expansion, but consider effects in
the radiation and matter periods of Standard Big Bang cosmology, assuming
a spectrum of gravitational waves produced in the early universe. We also
study the dependence of our results on the tensor to scalar ratio $r$ and
on the spectral index $n_T$.

What is new in our work concerning super-Hubble modes compared
to previous works is that we compute the magnitude and equation of 
state of the totality of IR modes. We consider the three phases
of the standard inflationary paradigm of cosmology, namely the
de Sitter phase, the radiation phase and the final matter phase,
and  study the dependence of the results on the tensor to scalar
ratio $r$ and on the tensor tilt $n_T$.
Regarding sub-Hubble modes, we consider the dependence of the results on $r$ and $n_T$
and discuss constraints on $r$ and $n_T$ which can be derived.

A word on our notation. We use the signature  $(-, +, +, +)$ for the metric. Physical
time is denoted $t$, and $x$ are the spatial comoving coordinates.
Greek letters are for space-time indices, Latin ones for space only
indices. The scale factor is denoted $a(t)$, and the Hubble expansion
rate by $H(t)$. It is often convenient to
work in terms of conformal time $\tau$ related to the physics time $t$
via $dt = a(t) d\tau$. In conformal time, the Hubble expansion rate
is denoted by ${\cal H}$. The derivative with respect to conformal
time is indicated by a prime.

\section{Effective Energy-Momentum Tensor of Gravitational Waves} 

We will consider a background space-time given by a homogeneous
and isotropic metric with scale factor $a(t)$.  To this background we add scalar and tensor
metric fluctuations of small amplitude which satisfy the linear fluctuation
equations (see e.g. \cite{MFB} for an overview of the theory of
cosmological perturbations and \cite{RHBrev} for a shorter introduction).
Because of the nonlinearities in the Einstein field equations, the
resulting metric fails to be a solution of the equations at second order.
In order to have a solution of the Einstein equations at second order, we
need to add second order terms in the metric. These include a correction
to the background metric and corrections to the fluctuations. Both of these
are of second order.

As described in \cite{Abramo1, Abramo2}, the correction to the background
metric can be found by inserting the ansatz for the metric with only linear
fluctuations into the Einstein equations, expanding to second order in the
amplitude of the fluctuations, and then taking the spatial average of the
resulting equations to extract the back-reaction effect on the background metric\footnote{
As described in \cite{Martineau}, the back-reaction on the fluctuation
mode with wavenumber $k$ can be determined similarly, with an integration
against $\exp (ikx)$ replacing the spatial averaging.
}. Each Fourier mode of
the fluctuations contributes independently to the back-reaction. 

The spatial averaging also eliminates any coupling between scalar and tensor metric
fluctuations at this order in perturbation theory. Hence, we can study
the back-reaction of scalar and tensor modes separately. Here we will focus
on the back-reaction of the tensor modes. Hence, we can set the scalar
fluctuations to zero and consider the metric
\begin{eqnarray}
ds^2 \, = \, -a^2 d\tau^2 +  a^2 \left[ \gamma_{ij} +  h_{ij}  \right] dx^i dx^j\, ,
\end{eqnarray} 
where $\gamma_{ij}$ is the spatial background metric, and $h_{ij}$ is the
transverse and traceless tensor of linearized gravitational waves.

The Einstein tensor at second order can be written as\footnote{
The Einstein tensor including scalar, vector and tensor perturbations up to second order is given 
in \cite{Bartolo:2004if} although the first-order vector and tensor perturbations are neglected there.
} 
\begin{eqnarray}
\delta\snd G^0_{~0}
&=&
\frac{1}{a^2} \left[ 
  \mathcal{H} h\sfst{km} h\sfst{'}_{km} 
+ \frac{1}{8} h\sfst{' km} h\sfst{'}_{km}
- \frac{1}{8} h\sfst{' k}_{~~k} h\sfst{'m}_{~~m}
- \frac{1}{2} h\sfst{km} \nabla^2 h_{km}
- \frac{1}{2} h\sfst{km} h\sfst{j}_{~~j,km}
\right.
\notag \\
&&~~
\left.
+ h\sfst{km} h\sfst{j}_{~~k,mj}
- \frac{1}{4} h\sfst{m~~,k}_{~~m} h\sfst{~~,j}_{kj}
+ \frac{1}{8} h\sfst{m~~,k}_{~~m} h\sfst{j}_{~j,k}
- \frac{3}{8} h\sfst{km,j} h_{km,j}
\right.
\notag \\
&&~~
\left.
+ \frac{1}{4} h\sfst{km}_{~~~~,k}
    \left( 2 h\sfst{~~~,j}_{mj}
	- h\sfst{j}_{~j,m}
	\right) 
+ \frac{1}{4} h\sfst{jk,m} h_{jm,k}
\right], \\
\delta\snd G^0_{~i} 
&=&
\frac{1}{a^2} \left[
 \frac{1}{2} h\sfst{mk}_{~~~~,m} h\sfst{'}_{ki} 
+ h\sfst{mk}  h\sfst{'}_{k[i,m]} 
- \frac{1}{4} h\sfst{' mk} h_{mk,i}
- \frac{1}{4} h\sfst{k~~,j}_{~~k}  h\sfst{'}_{ij}
\right], \\
\delta\snd G^i_{~0}
&=&
\frac{1}{a^2} \left[
  h\sfst{ij} h\sfst{' k}_{~~[k,j]}
+ \frac{1}{4} h\sfst{k}_{~~k,j} h\sfst{' ij}
- \frac{1}{2} h\sfst{~~~,j}_{jk}  h\sfst{'ik}
+ \frac{1}{4} h\sfst{'}_{jk} h\sfst{jk,i}
- h_{jk} h\sfst{'k[i,j]}
\right], \notag \\ \\
\delta\snd G^{i}_{~j}
&=&
\frac{1}{a^2} \delta^i_{~j} \left[ 
 \mathcal{H} h\sfst{'}_{km} h\sfst{km}
+ \frac{1}{2} h\sfst{''}_{km} h\sfst{km}
+ \frac{3}{8}h\sfst{'}_{km} h\sfst{' km} 
- \frac{1}{8} h\sfst{' k}_{~~~k} h\sfst{' m}_{~~~m} 
\right.
\notag \\
&&~~
\left.
- \frac{1}{2} h\sfst{km} 
  \left( \nabla^2 h_{km}
  	+ h\sfst{j}_{~j,km} 
          -2 h\sfst{j}_{~(k,m)j}
  \right)
- \frac{3}{8} h\sfst{km,j} h_{km,j}
\right.
\notag \\
&&~~
\left.
- \frac{1}{8} h\sfst{m~~,k}_{~~m} 
     \left( 2 h\sfst{,j}_{kj} 
     	- h\sfst{j}_{~~j,k}
	\right)
+ \frac{1}{4} h\sfst{km}_{~~~~,k} 
    \left( 2 h\sfst{~~~,j}_{mj} 
    - h\sfst{j}_{~~j,m}
    \right)
+ \frac{1}{4} h\sfst{jk,m} h_{jm,k}
\right]\notag \\
&& 
+
 \frac{1}{a^2} \left[
- \mathcal{H} h\sfst{ik} h\sfst{'}_{kj}
- \frac{1}{2} h\sfst{ik} h\sfst{''}_{kj}
- \frac{1}{2} h\sfst{' ik} h\sfst{'}_{kj}
+ \frac{1}{4} h\sfst{' k}_{~~k} h\sfst{' i}_{~~j}
+ \frac{1}{4} h\sfst{km,i} h_{km,j}
\right.
\notag \\
&&~~
\left.
+ \frac{1}{2} h\sfst{ik} h\sfst{m}_{~~m,kj}
+ \frac{1}{2} h\sfst{m~~,k}_{~~m} h\sfst{(i}_{~~k,j)}
- \frac{1}{4} h\sfst{m~~,k}_{~~m} h\sfst{i}_{~~j,k}
\right.
\notag \\
&&~~
\left.
+ \frac{1}{2} h\sfst{km} 
	\left( h\sfst{~~~,i}_{km~,j} 
	+ h\sfst{i}_{~~j,km} 
	- 2h\sfst{(i}_{~~m,j)k} 
	\right)
- \frac{1}{2} h\sfst{km}_{~~~,k} 
           \left( 2 h\sfst{~~(i}_{m~~,j)} 
	- h\sfst{i}_{~~j,m} 
	\right)
\right.
\notag \\
&&~~
\left.
+  h\sfst{i [k,m]} h_{kj,m} 
- h\sfst{ik} h\sfst{m}_{~~(k,j)m}
+ \frac{1}{2} h\sfst{ik} \nabla^2 h_{kj}
\right].
\end{eqnarray}
These expressions can be simplified by using the
transverse and traceless condition $\partial^i h_{ij} = h^i_{~i} =0$ on
the gravitational wave tensor. Then, the Einstein tensor simplifies to
\begin{eqnarray}
\delta\snd G^0_{~0}
&=&
\frac{1}{a^2} \left[ 
  \mathcal{H} h\sfst{km} h\sfst{'}_{km} 
+ \frac{1}{8} h\sfst{' km} h\sfst{'}_{km}
+\frac{1}{8}  h\sfst{km,j}  h_{km,j}
- \frac{1}{2}  \partial^j \left( h\sfst{km} h_{km,j} \right)
+ \frac{1}{4} \partial_k \left( h\sfst{jk,m} h_{jm} \right)
\right], \\
\delta\snd G^0_{~i} 
&=&
\frac{1}{a^2} \left[
h\sfst{mk}  h\sfst{'}_{k[i,m]} 
- \frac{1}{4} h\sfst{' mk} h_{mk,i}
\right], \\
\delta\snd G^i_{~0}
&=&
\frac{1}{a^2} \left[
 \frac{1}{4} h\sfst{'}_{jk} h\sfst{jk,i}
- h_{jk} h\sfst{'k[i,j]}
\right], \\
\delta\snd G^{i}_{~j}
&=&
\frac{1}{a^2} \delta^i_{~j} \left[ 
 \mathcal{H} h\sfst{'}_{km} h\sfst{km}
+ \frac{1}{2} h\sfst{''}_{km} h\sfst{km}
+ \frac{3}{8}h\sfst{'}_{km} h\sfst{' km} 
\right.
\notag \\
&&~~
\left.
- \frac{1}{2} \partial_j \left( h\sfst{km}   h_{km} \right)
+\frac{1}{2} h\sfst{km,j}   h_{km,j} 
- \frac{3}{8} h\sfst{km,j} h_{km,j}
+ \frac{1}{4} \partial^m \left( h\sfst{jk} h_{jm,k} \right)
\right]\notag \\
&& 
+
 \frac{1}{a^2} \left[
- \mathcal{H} h\sfst{ik} h\sfst{'}_{kj}
- \frac{1}{2} h\sfst{ik} h\sfst{''}_{kj}
- \frac{1}{2} h\sfst{' ik} h\sfst{'}_{kj}
+ \frac{1}{4} h\sfst{km,i} h_{km,j}
\right.
\notag \\
&&~~
\left.
+ \frac{1}{2} \partial_j \left( h\sfst{km} h\sfst{~~~,i}_{km~~} \right) 
- \frac{1}{2}  h\sfst{km}_{~~~,j}  h\sfst{~~~,i}_{km} 
+ \frac{1}{2} \partial_m \left( h\sfst{km} h\sfst{i}_{~j,k} \right)
\right.
\notag \\
&&~~
\left.
-  \partial_k \left( h\sfst{km} h\sfst{~~(i}_{m~~,j)} \right)
 - \frac12 \partial_m\left(  h\sfst{im,k} h_{kj} \right)
+ \frac{1}{2} \partial^m \left( h\sfst{ik}  h_{kj,m} \right)
\right]. 
\end{eqnarray}

The spatial average of a quantity $A$ is obtained by integrating over the 
constant time hypersurface and dividing by the spatial volume $V$ \cite{Abramo2}:
\begin{equation}
\llan A \rran \equiv \frac{1}{V} \lim_{V \rightarrow \infty} \int A dV.
\end{equation}
Note that in the presence of matter, this hypersurface has a physical meaning: it is
the surface of constant matter energy density.

Thus, denoting spatial averages by pointed parentheses, 
the spatially averaged Einstein tensors become
\begin{eqnarray}
\llan \delta\snd G^0_{~0} \rran
&=&
\frac{1}{a^2} \left[ 
  \mathcal{H} \llan h\sfst{km} h\sfst{'}_{km}  \rran
+ \frac{1}{8} \llan h\sfst{' km} h\sfst{'}_{km} \rran
+\frac{1}{8}  \llan h\sfst{km,j}  h_{km,j} \rran
\right], \\
\llan \delta\snd G^0_{~i}  \rran
&=&
\frac{1}{a^2} \left[
\llan h\sfst{mk}  h\sfst{'}_{k[i,m]}  \rran
- \frac{1}{4} \llan h\sfst{' mk} h_{mk,i} \rran
\right], \\
\llan \delta\snd G^i_{~0} \rran
&=&
\frac{1}{a^2} \left[
 \frac{1}{4} \llan h\sfst{'}_{jk} h\sfst{jk,i} \rran
- \llan h_{jk} h\sfst{'k[i,j]} \rran
\right], \\
\llan \delta\snd G^{i}_{~j} \rran
&=&
\frac{1}{a^2} \delta^i_{~j} \left[ 
 \frac{3}{8} \llan h\sfst{'}_{km} h\sfst{' km}  \rran
- \frac{3}{8} \llan h\sfst{km,n} h_{km,n} \rran
\right]\notag \\
&& 
+
 \frac{1}{a^2} \left[
- \frac{1}{2}  \llan h\sfst{' ik} h\sfst{'}_{kj} \rran
- \frac{1}{4} \llan h\sfst{km,i} h_{km,j} \rran
+ \frac{1}{2} \llan h\sfst{ik,m} h_{kj,m} \rran
\right],
\end{eqnarray}
where  the terms with a total derivative have been dropped since 
we are taking a spatial average.
In addition, we have also used  the equation of motion for $h_{ij}$:
\begin{equation}
\label{eq:EOM_chi}
h_{ij}^{''} + 2 \conh h_{ij}' - \nabla^2 h_{ij} = 0.
\end{equation}

Taking the above correction terms to the Einstein tensor to
the matter side of the cosmological equations, we can
read off the {\it effective energy density} and {\it effective pressure}
of gravitational waves. The energy density is
\begin{equation}
\rho_{\rm GW} =  \frac{1}{8\pi G} \delta G^0_{~0} = \frac{1}{8\pi G a^2} \left( \frac18 \llan (h'_{ij})^2 \rran + \frac18 \llan (\nabla h_{ij})^2  \rran 
+  \mathcal{H} \llan h^{ij} h{'}_{ij}  \rran \right),
\end{equation}
and the pressure is given by  \cite{Abramo2}
\begin{eqnarray}
p_{\rm GW} &=&\frac13 \frac{1}{8\pi G} G^i_{~i}  + \frac{1}{3 {\cal H}} \llan ^{(2)} \Gamma^{\alpha}_{~\alpha 0} \rran (\rho^{(0)} + p^{(0)} ) \notag \\
&=& \frac13 \frac{1}{8\pi G a^2} \left( - \frac58 \llan (h'_{ij})^2 \rran + \frac78 \llan (\nabla h_{ij})^2 \rran \right)
-  \frac{1}{3 {\cal H}} \llan ^{(2)} \Gamma^{\alpha}_{~\alpha 0} \rran (\rho^{(0)} + p^{(0)} ) \notag \\
&=& \frac13 \frac{1}{8\pi G a^2} \left( - \frac58 \llan (h'_{ij})^2 \rran + \frac78 \llan (\nabla h_{ij})^2 \rran \right)
+\frac12  \frac{1}{8\pi G a^2}   {\cal H} \left(1+ w^{(0)}  \right)  \llan h^{ij} h'_{ij} \rran \, .
\end{eqnarray}
Here $w^{(0)}$ is the equation of state parameter for the cosmological background 
(i.e., $  \rho^{(0)} + p^{(0)}  =  \rho^{(0)} (1+ w^{(0)} )$).
In the last equality, we have used the fact that 
\begin{equation}
\llan ^{(2)} \Gamma^{\alpha}_{~\alpha 0} \rran = -\frac12 \llan h^{ij} h'_{ij} \rran,
\end{equation}
and the Friedmann equation
\begin{equation}
H^2 = \frac{{\cal H}^2}{a^2} = \frac{8\pi G}{3} \rho^{(0)}.
\end{equation}

Now we work in Fourier space. One can expand $h_{ij} (\tau, {\bm x})$ as follows:
\begin{equation}
\label{eq:chi_ij_expand}
h_{ij}  (\tau, {\bm x}) = \sum_{\lambda = +, \times}  \int \frac{dk^3}{(2\pi)^3} \epsilon_{ij}^\lambda ({\bm k}) h_{\bm k}^{\lambda} (\tau) e^{i {\bm k}\cdot{\bm x}},
\end{equation}
where $ \epsilon_{ij}^\lambda$ is the polarization tensor which is symmetric in $i$ and $j$. 
It also satisfies the transverse-traceless condition and the normalization condition
\begin{equation}
\sum_{i,j} \llan  \epsilon_{ij}^\lambda ({\bm k})  \epsilon_{ij}^{\lambda'}   ({\bm k'}) \rran  \,
= \, 2 \delta^{\lambda \lambda'} \delta^{(3)}({\bm k} - {\bm k}').
\end{equation}
In Fourier space, the equation of motion for the gravitational waves is
\begin{equation}
\label{eq:EOM_h_k}
h^{''}_{\bm k} + 2 {\cal H} h^{'}_{\bm k}  + k^2 h_{\bm k} =0.
\end{equation}

By substituting the expansion given in Eq.~\eqref{eq:chi_ij_expand},  the energy density
and pressure of gravitational waves can be written as
\begin{eqnarray}
\rho_{\rm GW}^{\rm (tot)}  & = &  \frac{1}{8\pi G a^2}   \sum_\lambda  \int  \frac{dk}{k}   \frac{k^3}{\pi^2} \tilde{\rho}_{\rm GW} (k),   \\
p_{\rm GW}^{\rm (tot)} & = &  \frac{1}{8\pi G a^2}   \sum_\lambda  \int  \frac{dk}{k}   \frac{k^3}{\pi^2}  \tilde{p}_{\rm GW} (k),
\end{eqnarray}
where $\tilde{\rho}_{\rm GW} (k)$ and $\tilde{p}_{\rm GW} (k)$ are given by
\begin{eqnarray} \label{densities}
\tilde{\rho}_{\rm GW} (k)
& = &
 \frac18 \llan  |h_{\bm k}^{' \lambda}|^2 \rran + \frac18 k^2  \llan  |h_{\bm k}^{\lambda}|^2  \rran 
+  \mathcal{H} \llan  h_{\bm k}^{' \lambda}  h_{\bm k}^{\lambda \ast}   \rran,
  \\
\tilde{p}_{\rm GW}   (k)
& = &  
 - \frac{5}{24} \llan |h_{\bm k}^{' \lambda}|^2  \rran + \frac{7}{24} k^2 \llan  |h_{\bm k}^{\lambda}|^2 \rran
+\frac{3}{4}    {\cal H} \left(1+ w^{(0)}  \right)  \llan h_{\bm k}^{' \lambda}  h_{\bm k}^{\lambda \ast}   \rran.
\end{eqnarray}

Assuming that the initial amplitude of $h_{\bm k}^\lambda$ is $A_k^{\lambda} $, then at
any later time $h_{\bm k}^\lambda$ can be written as 
\begin{equation}
h_{\bm k}^\lambda \, = \, A_k^{\lambda} f(\tau, k) \, ,
\end{equation}
with $f(\tau, k)$ being the function which describes  the time evolution of $h_{\bm k}$
and equals to unity at the initial time.  The primordial GW spectrum (GW spectrum at the
initial time) is given by 
\begin{equation}
{\cal P}_{\rm prim} (k) = \sum_\lambda \frac{k^3}{\pi^2} \vert A_k^{\lambda} \vert^2 \, .
\end{equation}
Then, the total energy density $\rho_{\rm GW}^{\rm (tot)} $ and pressure $p_{\rm GW}^{\rm (tot)} $ 
in gravitational waves can be rewritten in terms of the primordial spectrum as
\begin{eqnarray}
\rho_{\rm GW}^{\rm (tot)} 
& = &
\frac{1}{8\pi G a^2}  \int  \frac{dk}{k} \frac{\tilde{\rho}_{\rm GW} (k)}{ \vert A_k^{\lambda} \vert^2}  {\cal P}_{\rm prim} (k),   \\
p_{\rm GW}^{\rm (tot)} 
& = &  
\frac{1}{8\pi G a^2}   \int  \frac{dk}{k}  \frac{\tilde{p}_{\rm GW} (k)}{ \vert A_k^{\lambda} \vert^2}   {\cal P}_{\rm prim} (k).
\end{eqnarray}
In the following we will assume that both polarization states have the same amplitude,
and we will drop the superscript $\lambda$ on the mode amplitude $A_k$.

As we will show in the next section,  for sub-Hubble modes the averages of the 
temporal and spatial gradient terms becomes the same and the terms proportional
to the Hubble parameter are negligible. 
In this case, the energy density $\rho_{\rm GW}$ and the pressure $p_{\rm GW}$ 
can be written as 
\begin{eqnarray}
\rho_{\rm GW} &=& \frac{1}{4 \cdot 8\pi G a^2}  \llan (\nabla h_{ij})^2 \rran, \\
p_{\rm GW} &=& \frac{1}{3 \cdot 4 \cdot 8\pi G a^2}  \llan (\nabla h_{ij})^2 \rran, 
\end{eqnarray}
which gives the expected equation of state
\begin{equation}
p_{\rm GW} = \frac13 \rho_{\rm GW}
\end{equation}
of radiation.

\section{Evolution of the Amplitude of Gravitational Waves}

Let us now briefly review the evolution of the amplitude $h_{\bm k}$
of the Fourier space gravitational wave modes. We consider three phases:
an initial de Sitter phase which generates the large-scale tensor modes,
a subsequent radiation (RD) phase and a final matter (MD) phase.

The equation of motion (EOM) for $h_{\bm k}$ is
\begin{equation}
\label{eq:EOM_chi_k}
h_{\bm k}^{''} + 2 \conh h_{\bm k}'  + k^2  h_{\bm k} = 0.
\end{equation}
Since $\conh$ is 
\begin{equation}
\label{HubbleE}
{\cal H} = -\frac{1}{\tau} ~~ ({\rm de~Sitter}),
\qquad 
{\cal H} = \frac{1}{\tau} ~~ ({\rm RD}),
\qquad
{\cal H} = \frac{2}{\tau} ~~ ({\rm MD}),
\end{equation}
one can introduce $x\equiv k \tau$ and Eq.~\eqref{eq:EOM_chi_k} can be rewritten as 
\begin{equation}
\frac{d^2 h_{\bm k}}{dx^2} + \frac{2A}{x} \frac{d h_{\bm k}}{dx}  + k^2  h_{\bm k} = 0,
\end{equation}
where $A = \tau \conh$.

We now look at the solutions in the three periods.
First, in the de Sitter period, the EOM is
\begin{equation}
\frac{d^2 h_{\bm k}}{dx^2} - \frac{2}{x} \frac{d h_{\bm k}}{dx}  + k^2  h_{\bm k} = 0,
\end{equation}
whose general solution is given by 
\begin{equation}
\label{eq:hk_dS}
h_{\bm k} = C_1 x \left( - \frac{\cos x}{x} - \sin x \right) + C_2 x \left( - \cos x + \frac{\sin x}{x} \right).
\end{equation}
Note that in the de Sitter phase $\tau$ is negative, and it tends towards $0$ at the end
of the phase, the time when we want to define the {\it primordial} spectrum. Thus we require 
that
\begin{equation}
\label{eq:BC}
h_{\bm k} \rightarrow A_k (={\rm const.})~~~ {\rm for}~~~x (=k\tau) \rightarrow 0.
\end{equation}
By expanding Eq.~\eqref{eq:hk_dS} around $x=0$, we get 
\begin{equation}
h = -C_1 -\frac12 C_1 x^2 + \frac16 C_2 x^3 + \cdots \, .
\end{equation}
To satisfy the condition Eq.~\eqref{eq:BC}, one should have $C_1 = - A_k$. Hence the 
long wavelength solution in de Sitter background can be written 
as 
\begin{equation}
h_{\bm k} = A_k \left( 1 + \frac12 (k\tau)^2 + \cdots \right), 
\end{equation}
which is the same as the one obtained in \cite{Abramo2}.

In the RD era the EOM becomes
\begin{equation}
\frac{d^2 h_{\bm k}}{dx^2} + \frac{2}{x} \frac{d h_{\bm k}}{dx}  + k^2  h_{\bm k} = 0 \, ,
\end{equation}
whose general solution is given by 
\begin{equation}
\label{eq:hk_RD}
h_{\bm k} = C_1 \frac{e^{- ix}}{x} + C_2   \frac{e^{ ix}}{x}.
\end{equation}
The small $x$ expansion of $h_{\bm k}$ is
\begin{equation}
h_{\bm k} = \frac{1}{x} \left( C_1+ C_2 \right) - i \left( C_1 - C_2 \right) + \frac12 \left( - C_1 -C_2 \right) x + \frac{i}{6} \left( C_1 - C_2 \right) x^2 + \cdots \, .
\end{equation}
We are only interested in modes which are super-Hubble at the end of the de Sitter phase (and
therefore at the beginning of the radiation phase). The initial conditions for $C_1$ and $C_2$
are hence given by taking the limit $x \rightarrow 0$ and equating with Eq.~\eqref{eq:BC}. 
We find
\begin{equation}
\label{eq:C_RD}
C_1 = - C_2 = \frac{i}{2} A_k.
\end{equation}
Therefore the long wavelength solution can be given by 
\begin{equation}
h_{\bm k} = A_k \left( 1 -  \frac16 (k\tau)^2 + \cdots \right), 
\end{equation}
which is again the same as the one obtained in \cite{Abramo2}.

The full solution in RD era can be given by (making use of Eq.~\eqref{eq:C_RD})
\begin{equation}
\label{eq:hk_RD_2}
h_{\bm k} = \frac{\sin x}{x} A_k,
\end{equation}
which can then also be used for the small wavelength region.

The EOM in the MD era takes the form
\begin{equation}
\frac{d^2 h_{\bm k}}{dx^2} + \frac{4}{x} \frac{d h_{\bm k}}{dx}  + k^2  h_{\bm k} = 0 \, ,
\end{equation}
and its general solution is 
\begin{equation}
\label{eq:hk_MD}
h_{\bm k} = \frac{C_1}{x^2} \left( - \frac{\cos x}{x} - \sin x \right) + \frac{C_2}{x^2} \left( - \cos x + \frac{\sin x}{x} \right) \, .
\end{equation}
The small $x$ expansion of Eq.~\eqref{eq:hk_MD} reads
\begin{equation}
h_{\bm k} = - \frac{C_1}{x^3} - \frac{C_1}{2x} + \frac{C_2}{3} + \frac{C_1}{8} x - \frac{C_2}{30} x^2 + \cdots \, .
\end{equation}
From condition Eq.~\eqref{eq:BC}, we can obtain 
\begin{equation}
C_1 = 0, ~~~ C_2 = 3 A_k \, ,
\end{equation}
and therefore the full solution is 
\begin{equation}
\label{eq:hk_MD_2}
h_{\bm k} = \frac{3 A_k}{x^2} \left( - \cos x + \frac{\sin x}{x} \right) \, .
\end{equation}
In the long wavelength limit, $h_{\bm k}$ is given by
\begin{equation}
h_{\bm k} = A_k \left( 1 -  \frac{1}{10} (k\tau)^2 + \cdots \right) \, .
\end{equation}
In the short wavelength limit where $k\tau \gg 1$, we have 
\begin{equation}
h_{\bm k} = -3 A_k\frac{\cos x}{x^2}.
\end{equation}

These equations will be used in the following to compute the back-reaction
effect of both short and long wavelength gravitational waves. At quadratic
order in the amplitude of fluctuations, each Fourier mode contributes independently
to the energy density and pressure of back-reaction. Hence, the contributions
of long (super-Hubble) and short (sub-Hubble) modes add up. We can write
the total energy density of back-reaction as
\begin{equation}
\label{eq:rho_GW_tot}
\rho_{\rm GW}^{\rm (tot)} 
=  \frac{1}{8\pi G a^2}  \left( \int_{k_{\rm min}}^{k_{\rm hor}} + \int_{k_{\rm hor}}^{k_{\rm max}} \right)  \frac{dk}{k}  \frac{\tilde{\rho}_{\rm GW} (k)}{ \vert A_k \vert^2} {\cal P}_{\rm prim} (k),
\end{equation}
where $k_{\rm hor} = a H$ corresponds to the Hubble radius at the time of Hubble
radius entry of the mode $k$.  $k_{\rm min}$ is an
infrared cutoff, and $k_{\rm max}$ an ultraviolet cutoff. 

If we have in mind fluctuations
which are generated during an early phase of inflation, then $k_{\rm max}$ should be taken
to correspond to the Hubble radius at the end of inflation (and hence is given by the energy 
scale at the end of inflation) since modes with $k > k_{\rm max}$ were always vacuum
fluctuations, were never squeezed and did not become classical\footnote{See e.g.
\cite{Kiefer, Martineau2} for a discussion of the classicalization of cosmological perturbations.}. 
Stated differently, the
effect of modes with $k > k_{\rm max}$ vanishes upon renormalization.

In the same context, one could consider the infrared cutoff to be given by the Hubble scale
at the beginning of inflation. This is a physical choice and it simply means that we will
not be considering the effects of modes for which the early inflationary phase makes no
predictions.

\section{Effects of Super-Hubble Modes}

For super-Hubble fluctuations, the kinetic energy contribution to the energy
density and pressure is of order $k^4$ and hence negligible. The contributions
of the spatial gradient term and the term proportional to ${\cal{H}}$ are both
of the same order of magnitude. Inserting the expressions for ${\cal{H}}$
from Eq.~(\ref{HubbleE}) and for the long wavelength limits of $h_k$ from the previous
section, one easily obtains
\begin{eqnarray}
\label{eq:chi_deSitter}
\tilde{\rho}_{\rm GW} (k)  & = & - \frac78 k^2 \llan \vert A_k^2 \vert   \rran ~~~({\rm de~Sitter}), \\
\tilde{\rho}_{\rm GW} (k)  & = & -  \frac{5}{24} k^2 \llan \vert A_k^2 \vert  \rran ~~~({\rm RD}), \\
\tilde{\rho}_{\rm GW} (k)  & = & -  \frac{11}{40} k^2 \llan \vert A_k^2 \vert \rran ~~~({\rm MD}).
\end{eqnarray}
In all of these cases, the pressure is given by 
\begin{equation}
\tilde{p}_{\rm GW} (k) = - \frac13 \tilde{\rho}_{\rm GW} (k) \, ,
\end{equation}
in agreement with what was derived in \cite{Abramo2}.

Note that the induced energy density of super-Hubble gravitational waves is
negative. This is true in all three cosmological phases. Also, in all three
phases the equation of state is that of curvature. Hence, we conclude
that the effects of gravitational waves lead to a change in the locally
measured energy density of spatial curvature. This contrasts to the
back-reaction effect of super-Hubble scalar cosmological perturbations
for which the equation of state is that of a negative cosmological constant
\cite{Abramo1, Abramo2}. For a discussion of the physical measurability
of this effect see e.g. \cite{Lam}. 

Let us now consider the effects of the totality of all super-Hubble modes during
the various phases. We will use the conventional notation for the power
spectrum of gravitational waves
\be \label{power-spectrum}
{\cal P}_{\rm prim} (k) = A_T (k_\ast) \left( \frac{k}{k_\ast} \right)^{n_T}  \, ,
\ee
where $k_\ast$ is the pivot scale at which the tensor spectrum is
normalized, $A_T (k_\ast)$ is the amplitude of the power spectrum at
that scale (in the following we will drop the argument on $A_T$),
and $n_T$ is the tensor spectral index.

To simplify the notation, we write
\be \label{IR-formula}
\tilde{\rho}_{\rm GW} (k)  \, = \,  - {\cal{C}} k^2 \llan \vert A_k^2 \vert   \rran \, 
\ee
where the constant ${\cal{C}}$ takes on various (positive) values in each
of the three phases we analyze (see Eq.~(\ref{eq:chi_deSitter})).

The general expression for the energy density in the totality of
super-Hubble modes is
\begin{equation}
\label{eq:rho_GW_IR}
\rho_{\rm GW}^{\rm (IR)} (t) \,
= \frac{1}{8\pi G a^2}  \int_{k_{\rm min}}^{k_{\rm hor}}  \frac{dk}{k}  
\frac{\tilde{\rho}_{\rm GW} (k)}{ \vert A_k^{\lambda} \vert^2} {\cal P}_{\rm prim} (k),
\end{equation}
Inserting the expressions Eqs.~(\ref{power-spectrum}) and (\ref{IR-formula})  into Eq.~(\ref{eq:rho_GW_IR})
we see that as long as $n_T > - 2$, the integral over $k$-modes gives 
\be
\rho_{\rm GW}^{\rm (IR)} (t) \, =  \, - \frac{k_{\rm{hor}}^2}{8 \pi G a^2} \frac{{\cal{C}} A_T }{2 + n_T} 
\left[ \left( \frac{k_{\rm{hor}}}{k_*} \right)^{n_T}  - \left( \frac{k_{\rm min}}{k_{\rm hor}} \right)^2 \left( \frac{k_{\rm min}}{k_\ast} \right)^{n_T} \right]
= -\frac{\rho_c (t)}{3}\frac{{\cal{C}} A_T }{2 + n_T} 
\left[ \left( \frac{k_{\rm{hor}}}{k_*} \right)^{n_T}  - \left( \frac{k_{\rm min}}{k_{\rm hor}} \right)^2 \left( \frac{k_{\rm min}}{k_\ast} \right)^{n_T} \right]
\, ,
\label{eq:rho_IR}
\ee
where we have used $ k_{\rm{hor}} \, = \, H (t) a(t)$ and the Friedmann equation. $\rho_c(t)$ is the critical background energy density at time $t$.
The value of $\rho_{\rm GW}^{\rm (IR)}$
is suppressed compared to the background energy density by the factor $A_T$
as should be expected since the back-reaction effect is quadratic in the amplitude
of the fluctuations and there is no secular growth as there is in the case of the 
back-reaction contribution of scalar metric fluctuations.

If the context of inflation with matter which satisfies the usual energy conditions,
the spectrum of gravitational waves must be red, i.e. $n_T < 0$. 
In this case, the value of $\rho_{\rm GW}^{\rm (IR)}$ increases as we go back in time.
Hence, from Eq.~(\ref{eq:rho_IR}) we can derive an upper bound on the number of $e$-foldings of
inflation. This comes from demanding that at the beginning of what we want the
period of inflation to be, the energy density in long wavelength gravitational waves
must be subdominant, i.e.
\be
\rho_{\rm GW}^{\rm (IR)}(t_i) \, \ll \frac{H^2(t_i)}{G} \, ,
\ee
or equivalently
\be \label{IRResult}
A_T \left( \frac{k_{\rm{hor}}(t_i)}{k_*} \right)^{n_T} \, \ll \, 1 \, ,
\ee
where $t_i$ is the initial time. Via the value of $A_T$, this bound will depend on
the tensor to scalar ratio. Since the amplitude of the tensor modes
is set in inflationary cosmology by the value of $H$, the above
requirement is essentially equivalent to demanding that at the
beginning of inflation $H$ is smaller than the Planck mass.

If the spectrum were blue (i.e. $n_T > 0$), as can be realized in Galileon inflation
\cite{Galinfl}, then the condition
\be
\rho_{\rm GW}^{\rm (IR)}(t_R) \, \ll \frac{H^2(t_R)}{G} \, ,
\ee
where $t_R$ corresponds to the end of inflation (reheating), sets an upper bound
on the value of $n_T$ (which, however, is similar to the requirement that the
fluctuations which exit the Hubble radius at the end of inflation are still in the
perturbative regime).

In both the radiation and matter phases, 
the general expression (\ref{eq:rho_GW_IR}) also yields
\be \label{RD_IR}
\rho_{\rm GW}^{\rm (IR)} \, \sim \, - A_T \rho_c(t) \left( \frac{k_{\rm{hor}}(t)}{k_*} \right)^{n_T} \, ,
\ee
where it is only the time dependence of $k_{\rm{hor}}(t)$ which is different. 
Note that in these phases, $k_{\rm{hor}}(t)$ is decreasing as time increases.
Hence, for $n_T < 0$ the relative contribution of gravitational waves to the energy
density increases in time (because modes with larger value of the gravitational
wave spectrum are entering the Hubble radius later). However, since the pivot
scale for the gravitational wave spectrum is typically taken to be within the linear
regime for cosmological perturbations, i.e. cosmological scale, the enhancement
factor $\bigl( \frac{k_{\rm{hor}}(t)}{k_*} \bigr)^{n_T}$ can never become large.

For $n_T > 0$, on the other hand, the relative importance of the effective
energy density of gravitational waves increases as we go back in time (again
because in this case it is at earlier times that modes with a higher amplitude
of the spectrum are entering the Hubble radius). There is then a constraint
on $n_T$ which reads:
\be \label{IRconstraint}
A_T  \left( \frac{k_{\rm{hor}}(t_R)}{k_*} \right)^{n_T} \, \ll \, 1 \, ,
\ee
where, as before $t_R$ is the beginning of the radiation phase. 
Note that
this constraint applies to all models of early universe cosmology, in
particular to the Ekpyrotic scenario \cite{Ekp} and to Pre-Big-Bang
cosmology \cite{PBB} which predict blue spectra of the tensor modes
with $n_T = 2$ in the case of the Ekpyrotic scenario and $n_T = 3$ in
the case of Pre-Big-Bang cosmology. String gas cosmology also
produces a blue spectrum \cite{BNPV2}, but with a very small spectral
index $n_T$.

Let us now study these constraints in more detail, referring back to Eq.~(\ref{eq:rho_IR}).
First we investigate the backreaction in de Sitter phase. 
During de Sitter phase, the range of $k_{\rm hor}$ is $ k_{\rm init} < k_{\rm hor} < k_R$ 
where $k_{\rm init}$ and $k_R$ are  the modes which exited the Hubble radius at the 
beginning and the end of inflation, respectively.  Here we assume instantaneous 
reheating and hence that the end of inflation  and the time of reheating are identical. Therefore 
we label the wavenumber corresponding to the end of inflation as $k_R$.
Assuming that the evolution of $H$ during inflation is described by\footnote{
Notice that  this formula is applicable for the slow-roll inflation model, and hence, 
strictly speaking, it cannot be  adopted to alternative scenarios like the Ekpyrotic model
\cite{Ekp}, Pre-Big-Bang cosmology \cite{PBB}, and string gas cosmology  \cite{BNPV2}.} 
\begin{equation}
\label{eq:EOM_Hubble}
n_T = - 2 \epsilon = 2 \frac{\dot{H}}{H^2}  = 2 \frac{1}{H} \frac{dH}{dN},
\end{equation}
$k_{\rm init} / k_\ast$ can be written as 
\begin{equation}
\label{eq:k_init}
\frac{k_{\rm init}}{ k_\ast} = \frac{a_{\rm init} H_{\rm init}}{a_\ast H_\ast} = \exp \left[ -\frac12 (2 + n_T) \left(N_{\rm tot}  - N_\ast \right) \right]
\end{equation}
where $H_\ast$ is the Hubble rate at the time when the mode $k_\ast$ exited the 
Hubble radius during inflation and $N_{\rm tot}$ is the total number of $e$-folds during 
inflation.  $N_\ast$ is the number of $e$-foldings counted backwards from the end of inflation to 
the time when the pivot scale $k_\ast$ exited the Hubble radius.
On the other hand,  the ratio $k_R / k_\ast$ can be given by 
\be
\label{eq:k_end_ast}
\frac{k_R}{k_\ast} = 
\frac{a_{\rm end} H_{\rm end}}{ a_\ast H_\ast} = e^{N_\ast} \frac{H_\ast e^{ n_T N_\ast /2}}{H_\ast}  
= \exp \left[ \frac12 (n_T + 2) N_\ast \right]. 
\ee

\begin{figure}[htbp]
\begin{center}
\resizebox{100mm}{!}{
     \includegraphics{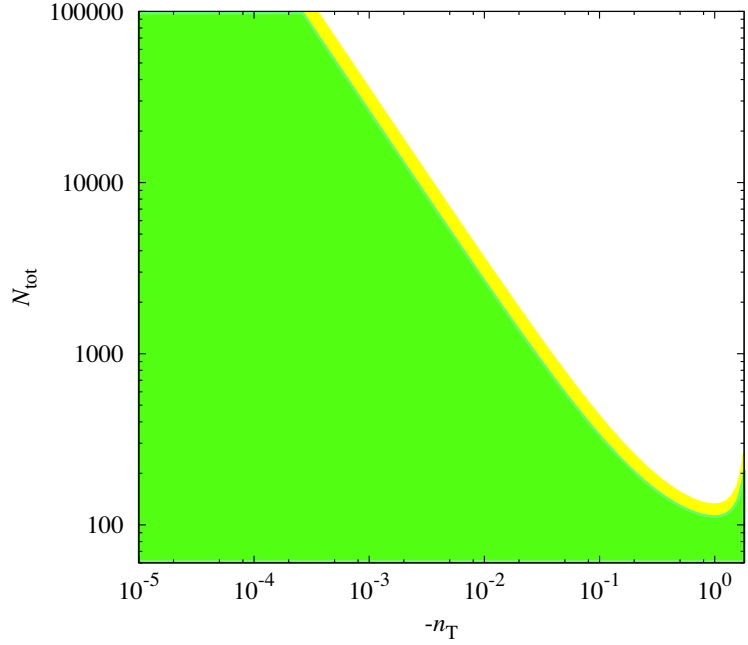}
}
\end{center}
\caption{Region with $|\rho_{\rm GW}^{\rm (IR)} | < \rho_c$ at $y=y_{\rm max}$ is shown for $r=10^{-6}$ (green) and $r=10^{-2}$ (yellow). $N_\ast = 60$ is assumed in this figure. The shaded regions are
the allowed ones.
}
\label{fig:Ntot_const}
\end{figure}

As shown in the following, we can obtain a bound for $N_{\rm tot}$ by requiring that 
$| \rho_{\rm GW}^{\rm (IR)} | < \rho_c $ when $n_T < 0$.
We rewrite Eq.~\eqref{eq:rho_IR}  as follows:
\be
\rho_{\rm GW}^{\rm (IR)} (t) \, = -\frac{\rho_c (t)}{3}\frac78 \frac{A_T }{2 + n_T} 
\left( \frac{k_{\rm min}}{k_\ast} \right)^{n_T} 
\left[ \left( \frac{k_{\rm{hor}}}{k_{\rm min}} \right)^{n_T}  - \left( \frac{k_{\rm min}}{k_{\rm hor}} \right)^2 \right]
\equiv
-\frac{\rho_c (t)}{3}\frac78 \frac{A_T }{2 + n_T}  
\left( \frac{k_{\rm min}}{k_\ast} \right)^{n_T}  f \left (y  \right) 
\, ,
\label{eq:rho_IR_dS}
\ee
where $y =k_{\rm hor} / k_{\rm min}$ with $k_{\rm min} = k_{\rm init}$ and  we have defined a function $f(y) = y^{n_T}  - y^{-2}$.
When $n_T < 0$,   $f(y)$ becomes the largest 
at  $y_{\rm max} = (-2 / n_T)^{1/(2 + n_T)} \left(=  (2 / |n_T|)^{1/(2 - |n_T|)}  \right)$ 
and its maximum value is given by
\be
f_{\rm max} (y_{\rm max}) = -\frac{1}{n_T} (2+n_T) \left( \frac{-n_T}{2} \right)^{2/(2+n_T)}
= \frac{1}{|n_T|} (2-|n_T|) \left( \frac{|n_T|}{2} \right)^{2/(2-|n_T|)}\, .
\ee
Since $ \left(k_{\rm min} / k_\ast \right)^{n_T}$  in Eq.~\eqref{eq:rho_IR_dS} depends on 
$N_{\rm tot}$ and gets larger 
as $N_{\rm tot}$ increases, we can obtain an upper bound on $N_{\rm tot}$ by 
requiring that $|\rho_{\rm GW}^{\rm (IR)} |$ does not 
exceed the background energy density $\rho_c$. In Fig.~\ref{fig:Ntot_const}, we show 
the region satisfying $|\rho_{\rm GW}^{\rm (IR)} | < \rho_c$
at $y = y_{\rm max}$ in the $n_T$--$N_{\rm tot}$ plane.  Interestingly, depending on 
the value of $n_T$, the total number of $e$-folds during inflation is constrained. 
The dependence of the total number of $e$-foldings of inflation is proportional
to $|n_T|^{-1}$ and to the logarithm of the amplitude of tensor modes.
As $n_T$ varies in the range $-10^{-1} \lesssim n_T \lesssim -10^{-2}$, 
the bound on $N_{\rm tot}$ increases from
$N_{\rm tot} = {\cal O}(300)$ to $ N_{\rm tot} = {\cal O}(3000)$.

Next we discuss the cases with RD and MD phases. 
In these phases,  $k_{\rm hor}$ is in the range  $k_{\rm eq} < k_{\rm hor} < k_R$ 
in the RD phase and $k_0 < k_{\rm hor} < k_{\rm eq}$ in the MD phase,
where $k_R$,  $k_{\rm eq}$ and  $k_0$ are the wavenumbers of the modes crossing the 
Hubble radius at the time of reheating,  the radiation-matter equality and 
the present time, respectively\footnote{
Here we neglect the dark energy dominated phase in the late Universe.
}. 
Therefore, in these phases, the contribution from the 2nd term in RHS of Eq.~\eqref{eq:rho_IR}  
is negligible compared to the 1st term since $k_{\rm min} \ll k_R, k_{\rm eq}, k_0$,
which allows us to approximate the result as 
\be
\rho_{\rm GW}^{\rm (IR)} (t) \, 
\simeq
-\frac{\rho_c (t)}{3}\frac{{\cal{C}} A_T }{2 + n_T} 
 \left( \frac{k_{\rm{hor}}}{k_*} \right)^{n_T} \, ,
\label{eq:rho_IR_approx}
\ee
For the case $n_T >0~ ( < 0) $, $\rho_{\rm GW}^{\rm (IR)} $ becomes maximal  
when $k_{\rm hor}$ takes the largest (smallest) possible value in the phase in 
consideration. In the RD phase, the maximum (minimum) value of $k_{\rm hor}$ is 
$k_{\rm hor} = k_R~ (k_{\rm eq})$. In the MD phase, the maximum (minimum) 
value  is  $k_{\rm hor} = k_{\rm eq}~(k_0) $. 

To evaluate $\rho_{\rm GW}^{\rm (IR)}$, we need to know 
that the ratios $ k_{\rm eq}/k_\ast$  and $k_0/k_\ast$  are given by
\begin{eqnarray}
\label{eq:k_eq_0_ast}
\frac{k_{\rm eq}}{k_\ast} 
&=&  \left( \frac{g_{\ast }(t_{\rm eq})}{g_{\ast}(t_R)} \right)^{1/2} \left( \frac{g_{\ast s}(t_R)}{g_{\ast s}(t_{\rm eq})} \right)^{1/3} \frac{T_{\rm eq}}{T_R} \, , \\
\frac{k_0}{k_\ast} 
&=&   \left( \frac{g_{\ast s}(t_R)}{g_{\ast s}(t_0)} \right)^{1/3} \frac{T_R}{T_0} \frac{H_0}{H_R} \, , 
\end{eqnarray}
where we have used entropy conservation, $g_\ast (t)$ is the effective number of 
degrees of freedom at time $t$,  and $g_{\ast s}(t)$ is its entropic counterpart. 
$T_{\rm eq}$ and $T_{\rm end}$ are given respectively by 
\begin{eqnarray}
\label{eq:T_eq}
&& T_{\rm eq} = \left( \frac{30 \rho_{\rm crit}}{\pi^2 g_{\ast} (t_{\rm eq}) } \Omega_r \right)^{1/4} \frac{\Omega_m}{\Omega_r} \, ,   \\
\label{eq:T_R}
&& T_{\rm R}= \left( \frac{90}{\pi^2 g_\ast (t_{\rm end})} \right)^{1/4} \sqrt{H_R M_{\rm pl}} \, .
\end{eqnarray}
The Hubble rate at the reheating $H_R$ can be related to $H_\ast$ by using
 Eq.~\eqref{eq:EOM_Hubble} and written as
\be
H_R = H_\ast \exp \left[ \frac{n_T}{2} N_\ast \right]\, .
\ee

\begin{figure}[htbp]
\begin{center}
\resizebox{100mm}{!}{
     \includegraphics{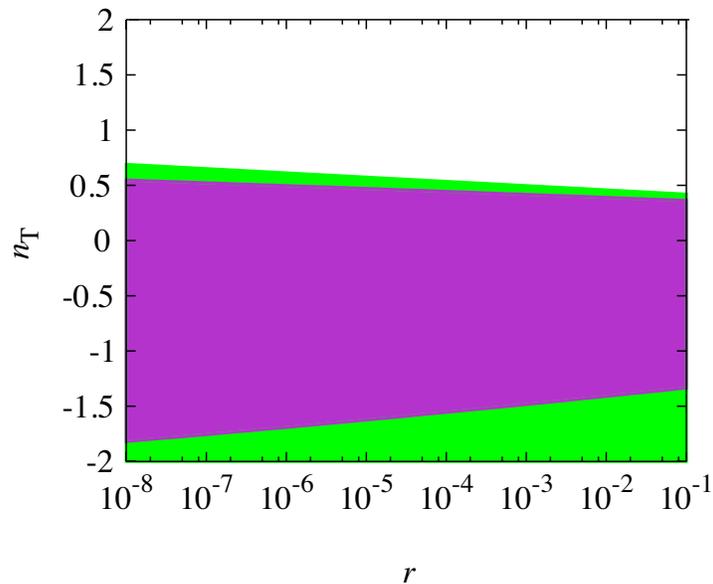}
}
\end{center}
\caption{The shaded regions in the $r$--$n_T$ plane are those allowed
from the constraint $ \left| \rho_{\rm GW}^{(IR)} \right|/\rho_c  < 1$  considering
super-Hubble fluctuations. The analysis is 
for the RD phase. The case assuming an evolution of $H$ during inflation according
to Eq.~\eqref{eq:EOM_Hubble}  is shown in purple. On the other hand, the case
where we assume that $H$ is constant during inflation is shown in green.
$N_\ast = 60$ is assumed in this figure. 
}
\label{fig:rho_GW_long_RD}
\end{figure}

By requiring that the backreaction $\rho_{\rm GW}^{(IR)}$ should not dominate over the 
background energy density $\rho_c (t)$,
some parameter regions in the $r$-$n_T$ plane can be excluded.
In Fig.~\ref{fig:rho_GW_long_RD}, we show the region satisfying 
$ \left| \rho_{\rm GW}^{(IR)} \right|  <\rho_c  $   in the $r$--$n_T$ plane
for the case of the RD phase. The purple and green regions are allowed given two 
different assumptions regarding the evolution of $H$ during inflation. 
The purple region in Fig.~\ref{fig:rho_GW_long_RD} is calculated by taking account of
the evolution of $H$ by adopting Eq.~\eqref{eq:EOM_Hubble}.
As already mentioned, Eq.~\eqref{eq:EOM_Hubble} may only be valid for 
slow-roll inflation model. Therefore we also consider the case where  $H$ is 
assumed to be constant during inflation, and the corresponding allowed region
is shown in green in the figure. As seen from the figure, when we assume that 
$H$ is constant, the constraint is less severe, especially in the region of red tilt. 
However, values of $n_T$  corresponding to a blue tilt are relatively well constrained 
in both of these cases.

The same argument also applies for the MD case. 
Its constraints on $r$ and $n_T$ are almost the same as the one obtained for the RD case, and 
hence we do not show the plot here. 
However, we note that, for the blue-tilted spectrum, the bound is not as severe compared to the RD case.
On the other hand, for values of $n_T < 0$ (red tilt), the constraint is slightly more stringent than 
in the RD case.

\section{Effects of Sub-Hubble Modes}

We now turn to the discussion of the effects of sub-Hubble modes. We
focus on the radiation and matter epochs. Note that in an early de Sitter phase
the sub-Hubble modes are in their vacuum state and hence yield
no back-reaction effects.

The general expression for the energy density in the totality of
sub-Hubble modes is
\begin{equation}
\label{eq:rho_GW_UV}
\rho_{\rm GW}^{\rm (UV)} \,
= \, \frac{1}{8\pi G a^2}  \int_{k_{\rm hor}}^{k_{\rm max}}  \frac{dk}{k}  
\frac{\rho_{\rm GW} (k)}{ \vert A_k \vert^2} {\cal P}_{\rm prim} (k),
\end{equation}
Inserting the expression (\ref{power-spectrum}) for the power spectrum
and the short wavelength solutions of the mode functions $h_k$ into
the general expression (\ref{densities}),  we obtain, for RD epoch
\ba
\rho_{\rm{GW}}^{\rm{(UV)}}(t) \, &=& \, \frac{1}{8 \pi G} \left( \frac{a_{\rm ref}}{a} \right)^4 H_{\rm ref}^2 
\frac{A_T(k_*)}{8} {\rm{log}} \left( \frac{k_{\rm{max}}}{k_{\rm{hor}}} \right) \,\,\, ({\rm{for}} \,\,\, n_T = 0)\,, \\ \notag \\
&=& \, \frac{1}{8 \pi G} \left( \frac{a_{\rm ref}}{a} \right)^4 H_{\rm ref}^2 \frac{1}{n_T} \frac{A_T(k_*)}{8} 
\left[ \left(\frac{k_{\rm{max}}}{k_*}  \right)^{n_T} - \left( \frac{k_{\rm{hor}}}{k_*} \right)^{n_T} \right]
\,\,\, ({\rm{for}} \,\,\, n_T \neq 0)\,,
\ea
where $a_{\rm ref}$ is the value of the scale factor at some reference time,
and $H_{\rm ref}$ is the Hubble rate at that time. Expressing the result in terms of the
critical energy density $\rho_c(t)$ this becomes
\ba \label{radresult}
\rho_{\rm{GW}}^{\rm{(UV)}}(t) \, 
&=& \, \frac{1}{3} \rho_c(t) 
\frac{A_T(k_*)}{8} {\rm{log}} \left( \frac{k_{\rm{max}}}{k_{\rm{hor}}} \right) \,\,\, ({\rm{for}} \,\,\, n_T = 0)\,, \\ \notag \\
&=& \, \frac{1}{3} \rho_c(t) \ \frac{A_T(k_*)}{8 n_T} 
\left[ \left(\frac{k_{\rm{max}}}{k_*}  \right)^{n_T} - \left( \frac{k_{\rm{hor}}}{k_*} \right)^{n_T} \right]
\,\,\, ({\rm{for}} \,\,\, n_T \neq 0) \,,
\ea
As expected, this energy density is positive and scales as radiation. For $n_T > 0$
the shortest wavelength modes dominate the contribution, and the relative contribution 
of gravitational waves to the energy density is (modulo subdominant terms) constant
in time. 
On the other hand, for $n_T < 0$,  the contribution of short wavelength
gravitational waves to the total energy density grows in time since
$k_{\rm{hor}}(t)$ is a decreasing function of time. 
In either case, $\rho_{\rm{GW}}^{\rm{(UV)}}$ takes the largest 
value for $k_{\rm max} = k_R$ and $k_{\rm hor} = k_{\rm eq}$.

\begin{figure}[htbp]
\begin{center}
\resizebox{100mm}{!}{
    \includegraphics{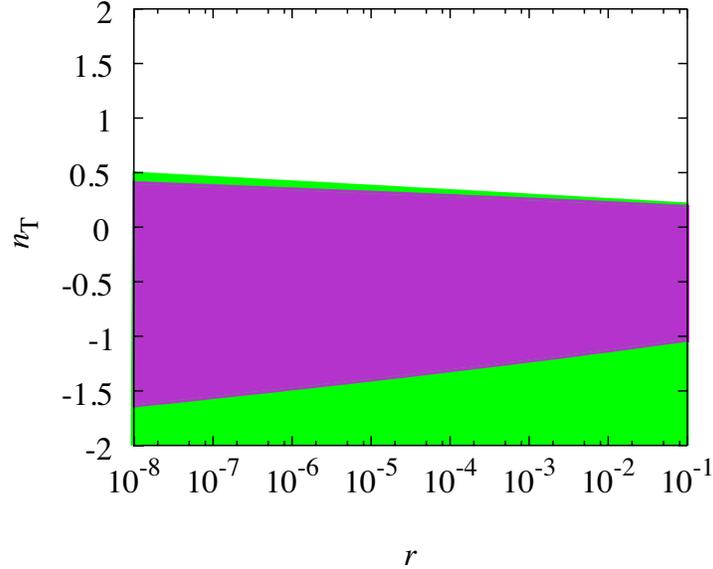}
}
\end{center}
\caption{The shaded regions in the $r$--$n_T$ plane are those allowed
from the constraint $ \Omega_{\rm GW} \left( = \left| \rho_{\rm GW}^{(IR)} \right|/\rho_c \right) < 1.5 \times 10^{-5}$  for short wavelength modes in the RD phase.
The case assuming an evolution of $H$ during inflation according
to Eq.~\eqref{eq:EOM_Hubble}  is shown in purple. On the other hand, the case
where we assume that $H$ is constant during inflation is shown in green.
$N_\ast = 60$ is assumed in this figure. 
}
\label{fig:rho_GW_short_RD}
\end{figure}

Short wavelength gravitational waves contribute to the effective number of degrees of
freedom of radiation. Nucleosynthesis constrains this number quite tightly
(see e.g. \cite{NSconstraints} for a review). The nucleosynthesis constraint
on the relative contribution $\Omega_{\rm{GW}}$ is\footnote{
The upper bound on $\Omega_{\rm GW}$ from the nucleosynthesis can be written  as \cite{NSconstraints}
\be
\Omega_{\rm GW} h^2 \le 5.6 \times10^{-6} (N_{\rm eff} - 3) \, ,
\ee
where $N_{\rm eff}$ is the upper bound on the effective number of neutrino species and $h$ is the Hubble constant 
in units of $100~{\rm km}/{\rm sec}/{\rm Mpc}$.
For a recent upper bound on $N_{\rm eff}$, see  \cite{Cyburt:2015mya}. 
Here we conservatively take $\Omega_{\rm GW} < 1.5 \times 10^{-5}$.
}
\be \label{GWbound}
\Omega_{\rm{GW}} \, < \, 1.5 \times 10^{-5} \, .
\ee
In Fig.~\ref{fig:rho_GW_short_RD}, we show the region of 
$\rho_{\rm{GW}}^{\rm{(UV)}} < 1.5 \times 10^{-5} \rho_c$
by taking $k_{\rm max} = k_R$ and $k_{\rm hor} = k_{\rm eq}$.
In particular, the bound on the tilt of the tensor spectrum becomes
\be
n_T \, < 0.2 \, 
\ee
if the tensor to scalar ratio is taken to be $r = 10^{-1}$. This is consistent with
the result in the first reference of \cite{Stewart}, where a value of $r \sim 1$ was used and the UV
cutoff was taken to be at the Planck scale.
Like in the case for super-Hubble modes, when the evolution of the Hubble parameter during inflation is taken into account by using Eq.~\eqref{eq:EOM_Hubble}, 
a red-tilted spectrum can also be constrained as $n_T \gtrsim  -1.1$ for $r \sim 0.01$.

As discussed in \cite{Stewart}, the current limits on the tilt $n_T$ from other
searches for gravitational waves such as direct detection experiments (LIGO)
or pulsar timing arrays yield weaker bounds on $n_T$ since these experiments
are less sensitive to the high frequency waves which dominate for positive
values of the tilt.

Next we look at the case in the MD phase.  Assuming that $n_T <2$, we obtain
\begin{eqnarray}
\label{eq:rho_short_MD}
 \rho_{\rm GW}^{\rm (UV)}
&=& 
 \frac{1}{8\pi G} \frac{a_R^6 H_R^4}{a^4 k_\ast^2}  \frac{9}{8 \cdot 16} \frac{1}{2-n_T}
\frac{ A_T(k_\ast)}{4}  \left[  \left( \frac{k_{\rm max}}{k_\ast} \right)^{n_T-2}  -  \left( \frac{k_{\rm hor}}{k_\ast} \right)^{n_T-2}  \right] \notag \\
&=& 
 \frac{\rho_c (t)}{3}  \frac{9}{8 \cdot 16} \frac{ A_T(k_\ast)}{2-n_T}    
  \left[ \left(  \frac{k_{\rm hor}}{k_\ast} \right)^{n_T} - \left(  \frac{k_{\rm hor}}{k_{\rm max}} \right)^2  \left(  \frac{k_{\rm max}}{k_\ast} \right)^{n_T} \right].
\end{eqnarray}
We can also constrain $n_T$ and $r$ by using the same argument as for the RD case.
In fact, for the MD case, when $n_T>0$, $ \rho_{\rm GW}^{\rm (UV)}$ gets largest for 
$k_{\rm hor} = k_{\rm eq}$. On the other hand,
when $n_T < 0$, the case of $k_{\rm hor} = k_0$ gives the maximum value of 
$ \rho_{\rm GW}^{\rm (UV)}$. Therefore, to get a constraint on $n_T$ and $r$,
we take $k_{\rm hor} = k_{\rm eq}$ and $k_0$ for $n_T >0$ and $<0$, respectively. 
Since the constraint on $n_T$ and $r$ from the MD case is similar to the one from the RD case, 
we do not show the plot here.

\section{Conclusions and Discussion}

In this paper we have re-visited the back-reaction effects of sub- and super-Hubble
modes of gravitational waves on the background cosmology. In agreement with
previous works \cite{Abramo2} we find that super-Hubble induce a {\bf negative}
energy density and have an equation of state which corresponds to curvature.
We find that if the tensor spectrum is blue (i.e. $n_T > 0$), then by demanding
that the energy density in the early universe is not dominated by  curvature leads
to some constraints. In the case of an inflationary universe, it leads to a constraint on the
total number of $e$-foldings which depends on the value of $n_T$ (see (\ref{IRResult})), 
and in the case of the radiation phase it leads to a constraint on the tensor index $n_T$
which depends on the energy scale at the beginning of the radiation phase 
(see (\ref{IRconstraint})). In both cases, the bounds depend on the tensor to 
scalar ratio at the pivot scale $k_*$.

The energy density in sub-Hubble gravitational waves (which behave like
normal radiation) is constrained by the nucleosynthesis bound (\ref{GWbound}).
If the tensor spectrum is blue, this leads to constraints of the gravitational
wave spectrum parameters $r$ and $n_T$ which are given by Eq.~(\ref{radresult})
and depicted graphically in Figure 3. 
When the evolution of the Hubble parameter during inflation is taken into account (see Eq.~\eqref{eq:EOM_Hubble}), which is the case 
for usual slow-roll inflationary models, we can also obtain a lower bound on $n_T$ from the the above argument (see Figs.~\ref{fig:rho_GW_long_RD} and \ref{fig:rho_GW_short_RD}).

\section*{Acknowledgement}
\noindent

T.T would like to thank Cosmology group at McGill University for the hospitality during the visit, where a part of this work was done.
The research at McGill is supported in
part by funds from NSERC and from the Canada Research Chair program.
This work is partially supported by JSPS KAKENHI Grant Number 15K05084  (TT), 17H01131 (TT), and MEXT KAKENHI Grant Number 15H05888 (TT).


\begin{thebibliography}{99}

\bibitem{Starob}
A.~A.~Starobinsky,
  ``Spectrum of relict gravitational radiation and the early state of the universe,''
  JETP Lett.\  {\bf 30} (1979) 682
   [Pisma Zh.\ Eksp.\ Teor.\ Fiz.\  {\bf 30} (1979) 719].
  
\bibitem{infl}
A. Guth, ``The Inflationary Universe: A Possible Solution
To The Horizon And Flatness Problems,'' Phys.\ Rev.\  D {\bf 23}, 347
(1981);\\ 
R.~Brout, F.~Englert and E.~Gunzig, ``The Creation Of The Universe As A
Quantum Phenomenon,'' Annals Phys.\  {\bf 115}, 78 (1978);\\ 
A.~A.~Starobinsky, ``A New Type Of Isotropic Cosmological Models Without
Singularity,'' Phys.\ Lett.\ B {\bf 91}, 99 (1980);\\ 
K.~Sato, ``First Order Phase Transition Of A Vacuum And Expansion Of The
Universe,'' Mon.\ Not.\ Roy.\ Astron.\ Soc.\  {\bf 195}, 467 (1981);\\
L.~Z.~Fang, ``Entropy Generation in the Early Universe by Dissipative
Processes Near the Higgs' Phase Transitions,'' Phys.\ Lett.\ B {\bf 95},
154 (1980). 

\bibitem{BNPV2}
R.~H.~Brandenberger, A.~Nayeri, S.~P.~Patil and C.~Vafa,
  ``Tensor Modes from a Primordial Hagedorn Phase of String Cosmology,''
  Phys.\ Rev.\ Lett.\  {\bf 98}, 231302 (2007)
  [hep-th/0604126];\\
R.~H.~Brandenberger, A.~Nayeri and S.~P.~Patil,
  ``Closed String Thermodynamics and a Blue Tensor Spectrum,''
  Phys.\ Rev.\ D {\bf 90}, no. 6, 067301 (2014)
  [arXiv:1403.4927 [astro-ph.CO]].
  
\bibitem{BV}
R.~H.~Brandenberger and C.~Vafa, 
``Superstrings In The Early Universe,'' 
Nucl.\ Phys.\ B {\bf 316}, 391 (1989). 

\bibitem{SGrev}
R.~H.~Brandenberger, ``String Gas Cosmology: Progress
and Problems,'' Class.\ Quant.\ Grav.\  {\bf 28}, 204005 (2011)
[arXiv:1105.3247 [hep-th]];\\
R.~H.~Brandenberger, ``String Gas Cosmology,'' String Cosmology,
J.Erdmenger (Editor).  Wiley, 2009. p.193-230 [arXiv:0808.0746
[hep-th]];\\ 
 T.~Battefeld and S.~Watson, ``String gas cosmology,'' Rev.\ Mod.\
 Phys.\  {\bf 78}, 435 (2006) [arXiv:hep-th/0510022]. 
 
\bibitem{BICEP}
  P.~A.~R.~Ade {\it et al.} [BICEP2 and Keck Array Collaborations],
  ``Improved Constraints on Cosmology and Foregrounds from BICEP2 and Keck Array Cosmic Microwave Background Data with Inclusion of 95 GHz Band,''
  Phys.\ Rev.\ Lett.\  {\bf 116}, 031302 (2016)
  [arXiv:1510.09217 [astro-ph.CO]].

  
\bibitem{Stewart}
A.~Stewart and R.~Brandenberger,
  ``Observational Constraints on Theories with a Blue Spectrum of Tensor Modes,''
  JCAP {\bf 0808}, 012 (2008)
  [arXiv:0711.4602 [astro-ph]];\\
R.~Camerini, R.~Durrer, A.~Melchiorri, A.~Riotto, R.~Durrer, A.~Melchiorri and A.~Riotto,
  ``Is cosmology compatible with blue gravity waves?,''
  Phys.\ Rev.\ D {\bf 77}, 101301 (2008)
  [arXiv:0802.1442 [astro-ph]];\\
 S.~Kuroyanagi, T.~Takahashi and S.~Yokoyama,
  ``Blue-tilted Tensor Spectrum and Thermal History of the Universe,''
  JCAP {\bf 1502}, 003 (2015)
  [arXiv:1407.4785 [astro-ph.CO]];\\
  G.~Cabass, L.~Pagano, L.~Salvati, M.~Gerbino, E.~Giusarma and A.~Melchiorri,
  ``Updated Constraints and Forecasts on Primordial Tensor Modes,''
  Phys.\ Rev.\ D {\bf 93}, no. 6, 063508 (2016)
  [arXiv:1511.05146 [astro-ph.CO]].
     
\bibitem{Wald}
S.~R.~Green and R.~M.~Wald,
  ``How well is our universe described by an FLRW model?,''
  Class.\ Quant.\ Grav.\  {\bf 31}, 234003 (2014)
  [arXiv:1407.8084 [gr-qc]].
  
\bibitem{Fabio1}
F.~Finelli and R.~H.~Brandenberger,
  ``Parametric amplification of gravitational fluctuations during reheating,''
  Phys.\ Rev.\ Lett.\  {\bf 82}, 1362 (1999)
  [hep-ph/9809490].
  
\bibitem{Abramo1}
V.~F.~Mukhanov, L.~R.~W.~Abramo and R.~H.~Brandenberger,
  ``On the Back reaction problem for gravitational perturbations,''
  Phys.\ Rev.\ Lett.\  {\bf 78}, 1624 (1997)
  [gr-qc/9609026].
  
\bibitem{Abramo2}
L.~R.~W.~Abramo, R.~H.~Brandenberger and V.~F.~Mukhanov,
  ``The Energy - momentum tensor for cosmological perturbations,''
  Phys.\ Rev.\ D {\bf 56}, 3248 (1997)
  [gr-qc/9704037].
  
\bibitem{RHBbrReview}
R.~H.~Brandenberger,
  ``Back reaction of cosmological perturbations and the cosmological constant problem,''
  hep-th/0210165.
  
\bibitem{Polyakov}
A.~M.~Polyakov,
  ``Infrared instability of the de Sitter space,''
  arXiv:1209.4135 [hep-th];\\
  A.~M.~Polyakov,
  ``Decay of Vacuum Energy,''
  Nucl.\ Phys.\ B {\bf 834}, 316 (2010)
  [arXiv:0912.5503 [hep-th]];\\
  A.~M.~Polyakov,
  ``De Sitter space and eternity,''
  Nucl.\ Phys.\ B {\bf 797}, 199 (2008)
  [arXiv:0709.2899 [hep-th]].
  
\bibitem{WT1}
N.~C.~Tsamis and R.~P.~Woodard,
  ``Relaxing the cosmological constant,''
  Phys.\ Lett.\ B {\bf 301}, 351 (1993); \\
N.~C.~Tsamis and R.~P.~Woodard,
  ``Quantum gravity slows inflation,''
  Nucl.\ Phys.\ B {\bf 474}, 235 (1996)
  [hep-ph/9602315];\\
 N.~C.~Tsamis and R.~P.~Woodard,
  ``Comment on `Can infrared gravitons screen Lambda?',''
  Phys.\ Rev.\ D {\bf 78}, 028501 (2008)
  [arXiv:0708.2004 [hep-th]].
     
\bibitem{WT}
N.~C.~Tsamis and R.~P.~Woodard,
  ``The Physical basis for infrared divergences in inflationary quantum gravity,''
  Class.\ Quant.\ Grav.\  {\bf 11}, 2969 (1994);\\
N.~C.~Tsamis and R.~P.~Woodard,
  ``The Quantum gravitational back reaction on inflation,''
  Annals Phys.\  {\bf 253}, 1 (1997)
  [hep-ph/9602316].
  
\bibitem{Unruh}
W.~Unruh,
  ``Cosmological long wavelength perturbations,''
  astro-ph/9802323.
  
\bibitem{Ghazal1}
G.~Geshnizjani and R.~Brandenberger,
  ``Back reaction and local cosmological expansion rate,''
  Phys.\ Rev.\ D {\bf 66}, 123507 (2002)
  [gr-qc/0204074].
  
\bibitem{AW}  
L.~R.~Abramo and R.~P.~Woodard,
  ``No one loop back reaction in chaotic inflation,''
  Phys.\ Rev.\ D {\bf 65}, 063515 (2002)
  [astro-ph/0109272].
  
\bibitem{Ghazal2}
G.~Geshnizjani and R.~Brandenberger,
  ``Back reaction of perturbations in two scalar field inflationary models,''
  JCAP {\bf 0504}, 006 (2005)
  [hep-th/0310265].
  
\bibitem{Vacca}
G.~Marozzi, G.~P.~Vacca and R.~H.~Brandenberger,
  ``Cosmological Backreaction for a Test Field Observer in a Chaotic Inflationary Model,''
  JCAP {\bf 1302}, 027 (2013)
  [arXiv:1212.6029 [hep-th]].
  
 \bibitem{Lam}
 R.~H.~Brandenberger and C.~S.~Lam,
  ``Back-reaction of cosmological perturbations in the infinite wavelength approximation,''
  hep-th/0407048.
  
\bibitem{MFB}
V.~F.~Mukhanov, H.~A.~Feldman and R.~H.~Brandenberger,
``Theory of cosmological perturbations. Part 1. Classical perturbations.
Part 2. Quantum theory of perturbations. Part 3. Extensions,'' Phys.\
Rept.\  {\bf 215}, 203 (1992). 

\bibitem{RHBrev}
R.~H.~Brandenberger, ``Lectures on the theory of
cosmological perturbations,'' Lect.\ Notes Phys.\  {\bf 646}, 127 (2004)
[arXiv:hep-th/0306071]. 

\bibitem{Martineau} 
  P.~Martineau and R.~H.~Brandenberger,
  ``The Effects of gravitational back-reaction on cosmological perturbations,''
  Phys.\ Rev.\ D {\bf 72}, 023507 (2005)
  [astro-ph/0505236].

\bibitem{Bartolo:2004if} 
  N.~Bartolo, E.~Komatsu, S.~Matarrese and A.~Riotto,
  ``Non-Gaussianity from inflation: Theory and observations,''
  Phys.\ Rept.\  {\bf 402}, 103 (2004)
  [astro-ph/0406398].
  
\bibitem{Kiefer}
C.~Kiefer, D.~Polarski and A.~A.~Starobinsky,
  ``Quantum to classical transition for fluctuations in the early universe,''
  Int.\ J.\ Mod.\ Phys.\ D {\bf 7}, 455 (1998)
  [gr-qc/9802003].
  
\bibitem{Martineau2}
P.~Martineau,
  ``On the decoherence of primordial fluctuations during inflation,''
  Class.\ Quant.\ Grav.\  {\bf 24}, 5817 (2007)
  [astro-ph/0601134].
  
\bibitem{Galinfl}
T.~Kobayashi, M.~Yamaguchi and J.~Yokoyama,
  ``G-inflation: Inflation driven by the Galileon field,''
  Phys.\ Rev.\ Lett.\  {\bf 105}, 231302 (2010)
  [arXiv:1008.0603 [hep-th]].
  
\bibitem{Ekp}
J.~Khoury, B.~A.~Ovrut, P.~J.~Steinhardt and N.~Turok,
``The Ekpyrotic universe: Colliding branes and the origin of the hot big
bang,'' 
Phys.\ Rev.\ D {\bf 64}, 123522 (2001) [hep-th/0103239].

\bibitem{PBB}
M.~Gasperini and G.~Veneziano, 
``Pre - big bang in string cosmology,'' 
Astropart.\ Phys.\  {\bf 1}, 317 (1993)
[hep-th/9211021]. 

\bibitem{NSconstraints}
M.~Maggiore,
  ``Gravitational wave experiments and early universe cosmology,''
  Phys.\ Rept.\  {\bf 331}, 283 (2000)
  [gr-qc/9909001].
  
 
 \bibitem{Cyburt:2015mya} 
  R.~H.~Cyburt, B.~D.~Fields, K.~A.~Olive and T.~H.~Yeh,
  ``Big Bang Nucleosynthesis: 2015,''
  Rev.\ Mod.\ Phys.\  {\bf 88}, 015004 (2016)
  [arXiv:1505.01076 [astro-ph.CO]]. 
  
  
\end{thebibliography}
\end{document}